\documentclass[10pt,5p,times]{elsarticle}

\usepackage{amssymb}
\usepackage{lipsum} 
\usepackage{slashed}
\usepackage{amsmath}
\usepackage{amsfonts}
\usepackage{xcolor}
\usepackage{placeins}
\newcommand{\nopieft}{\mbox{$\slashed{\pi}$}EFT}

\newcommand{\be}{\begin{equation}}
\newcommand{\ee}{\end{equation}}

\newcommand{\bea}{\begin{eqnarray}}
\newcommand{\eea}{\end{eqnarray}}

\newcommand{\AS}{\hat{\mathcal{A}}}

\newcommand{\Avec}{\ensuremath{\boldsymbol{A}}}

\newcommand{\rvec}{\ensuremath{\boldsymbol{r}}}

\newcommand{\xvec}{\ensuremath{\boldsymbol{x}}}

\journal{Physics Letters B}
\begin{document}

\begin{frontmatter}

\title{Five-body calculation of $s$-wave $n$-$^4$He scattering at next-to-leading order \nopieft}

\author[heb]{Mirko Bagnarol}
\author[heb,npi]{Martin Sch\"afer}
\ead{m.schafer@ujf.cas.cz}
\author[heb]{Betzalel Bazak}
\author[heb]{Nir Barnea}

\affiliation[heb]{organization={The Racah Institute of Physics},
            addressline={The Hebrew University}, 
            city={Jerusalem},
            postcode={9190401}, 
            country={Israel}}

\affiliation[npi]{organization={Nuclear Physics Institute of the Czech Academy of Sciences},
            city={Rez},
            postcode={25068}, 
            country={Czech Republic}}

\begin{abstract}
We present the first five-body calculations of $s$-wave $n$-$^4$He scattering within leading order and next-to-leading order (NLO) pionless effective field theory (\nopieft). Using an harmonic oscillator trap technique and \nopieft\ fitted to just six well-established experimental parameters, we predict the $s$-wave $n$-$^4$He phase shifts, scattering length $a^{1/2}_{n ^4\text{He}}(\text{NLO})=2.47(4\ \text{num.})~(17\ \text{theor.})~{\rm fm}$, and effective range $r^{1/2}_{n ^4\text{He}}(\text{NLO})=1.384(3\ \text{num.})~(211\ \text{theor.})~{\rm fm}$ in agreement with experiment. The apparent cutoff independence of our results is used to estimate the theoretical errors coming as an integral part of our final results.
\end{abstract}



%\begin{keyword}


%keyword one \sep keyword two




%\end{keyword}

\end{frontmatter}


%=======================================
\section{Introduction} \label{sec:intro}
%=======================================

Few-body scattering and reactions play an important role in our understanding of  the nuclear force and in testing its different theoretical representations available at hand. Recent advances in computer power and computational techniques have allowed to address relevant observables with unprecedented precision. With controlled contribution from numerical uncertainties or from a specific few-body approach, this widens up the possibility for a quantitative assessment of different theoretical models of nuclear interaction.

Low-momentum $n-{\rm ^4He}$ elastic scattering is a subject of a lasting interest. With only one spin-doublet channel in $s$-wave, there is enough high quality experimental data regarding the scattering length $a_{n {\rm ^4He}}^{1/2}$ \cite{GBSC63,REA69,KRBB79,HWAH20}, allowing for a detailed and meaningful comparison with theory. In particular, a recent neutron interferometric datum $a_{n {\rm ^4He}}^{1/2}=2.4746(17\ {\rm stat.})(11\ {\rm syst.})~{\rm fm}$ \cite{HWAH20} gives a very stringent constraint. At non-zero momenta and in higher partial waves, scattering information is available from a series of $R$-matrix studies \cite{HB66,SW72,AR73,BF77,Hale08}. These works clearly indicate that elastic $^2 P_{3/2}$ and $^2 P_{1/2}$ phase shifts below the $\rm ^4 He$ breakup threshold are strongly affected by two low-laying and relatively narrow resonances ${\rm ^5He}~(J^\pi = 3/2^-)$ and ${\rm ^5He}~(J^\pi = 1/2^-)$. Furthermore, the contribution from higher $l>1$ partial waves to the total cross-section seems to be considerably smaller.

From the numerical perspective, the calculation of five-body scattering is still a challenging task. The first microscopic study of $n-{\rm ^4He}$ scattering using realistic interactions was performed more than a generation ago within the Monte Carlo approach \cite{CSK87}, see Ref. \cite{NPWC07} for a more recent work. $n-{\rm ^4He}$ scattering was also addressed by means of the Faddeev-Yakubovsky (FY) formalism with various types of nuclear potentials \cite{Lazauskas18,Lazauskas20,LC20}. It was further considered as a testing ground to assess different few-body techniques such as no-core shell-model with the resonating group method (NCSM/RGM) \cite{QN08,QN09,NRQ10,HLNQ13}, no-core shell model with continuum (NCSMC) \cite{NQHRR16,ZSNG20,KQQW20}, or NCSM with a confining potential trap \cite{ZSNG20} using different Chiral effective field theory ($\chi$EFT) interactions. Finally, the single-state harmonic oscillator representation of scattering equation method (SS-HORSE) \cite{SMMV16,SMMM18} or symmetry-adapted NCSM/RGM \cite{MLDE22} were applied to address the $n-{\rm ^4He}$ scattering as well. 

Most of the works employing $\chi$EFT forces can describe the global characteristics of $s$-wave scattering reasonably well, but fail to reproduce the experimental value of the scattering length. In addition, they tend to underestimate the splitting between the 
$^2 P_{3/2}$ and the $^2 P_{1/2}$ phase shifts. In fact, recent analysis of 
the theoretical uncertainties by Kravvaris \emph{et al.} \cite{KQQW20} showed, by varying 
low energy constants (LECs) of the $\chi$EFT($\rm N^2LO$) three-body force, that it is not possible to describe satisfactorily the $^2 P_{3/2}$ phase shifts using the $\chi$EFT($\rm N^3LO$) \cite{EM03} or $\chi$EFT($\rm N^4LO$) \cite{EMN17}  two-body forces.

Pionless effective field theory (\nopieft) \cite{HKK20}, presents a different approach for describing low-energy nuclear interaction. In this theory, pionic degrees of freedom are integrated out leaving the neutrons and the protons as the only relevant degrees of freedom. Consequently, the breakdown scale of the theory is given by the pion mass $m _{\pi}$. For few-body nuclear systems with a relatively small typical momentum $Q$, the nuclear interaction is predominantly driven by the relatively large $NN$ scattering lengths (with respect to the range of the nuclear interaction $R \approx \hbar/m_\pi c \simeq 1.4$~fm). Therefore, these systems belong to a universality class characterized by large scattering lengths, and their properties can be described by a rather small set of experimental constraints.

\nopieft\ describes the nuclear interaction by reproducing the effective range expansion (ERE) \cite{Kolck99}. At leading order (LO), a three-body scale must be included introducing a contact three-body force \cite{BedHamKol99}. Furthermore, at next-to-leading order (NLO) also a contact four-body force needs to be included \cite{SB23,BazKirKon19}. Finally, a momentum-dependent three-body force enters at next-to-next-to-leading order (N$^2$LO) \cite{BedRupGri03,Gri04,Gri05}. Currently, the status of momentum-dependent four-body force or five- and higher-body interactions is not clear.

Considering \nopieft\ studies of $n$-nucleus reactions, $n$-$^2$H scattering was successfully addressed in multiple works \cite{BedKol98,BedHamKol98,GabBedGri00,BedGri00,HamMeh01,BedRupGri03,Gri04,Van13,MarSprVan16,Kon17,RupVagHig19} up to N$^2$LO. The low-energy $s$-wave $n$-$^3$H and $n$-$^3$He scattering were calculated at LO in Ref.~\cite{Kir13} and up to NLO in Refs.~\cite{KirGriShu10,SB23}, where in the latter a reassuring agreement with other theoretical results or relevant experimental data was achieved. So far, the $n$-$^4$He elastic scattering has only been  studied within the confines of an EFT with explicit neutron and $\alpha$-particle degrees of freedom \cite{BHK02}.   

In this letter, we present the first \nopieft\ five-body calculation of $s$-wave $n-{\rm ^4He}$ scattering with perturbative inclusion of NLO terms. The structure of the letter is as follows. In Sec. \ref{sec:mod} we describe the $\slashed{\pi}$EFT potential at LO and NLO. In Sec. \ref{sec:meth} we describe the numerical methods employed to solve the five-body Schr\"odinger equation and to extract the $n-{\rm ^4He}$ phase shifts. Our results are shown and compared to other theoretical predictions and experimental values in Sec \ref{sec:res}. Sec. \ref{sec:con} concludes our work.

%=============================
\section{Model}\label{sec:mod}
%=============================

%=========================== Leading order
We use \nopieft\ up to NLO to describe the nuclear interaction. At LO, the \nopieft\ potential consists of two and three-body contact terms acting in the relevant $s$-wave channels. To solve the Schr\"{o}dinger equation, singular terms in the potential need to be regularized. We use a local Gaussian regulator $\delta _\Lambda (\rvec_{ij}) = e^{-\frac{ \Lambda ^2}{4} \rvec^2_{ij}}$ with a momentum cutoff $\Lambda$ smearing the LO contact interaction over a distance $\Lambda^{-1}$. Here,  $\rvec_{ij} = \rvec_i - \rvec_j$ is the distance between nucleons $i$ and $j$. After the regularization, the LO potential possesses the form

\begin{align}\label{eq:V_LO}
  V^{(0)} &= \sum _{i<j} \Big( C_0^{(0)}(\Lambda) \hat P _{ij}^{(1,0)} 
                 + C_1^{(0)}(\Lambda) \hat P_{ij}^{(0,1)} \Big) 
                 ~\delta_\Lambda(\rvec_{ij}) + 
  \cr & 
  + \sum _{i<j<k} \sum _{\text{cyc}} 
  D_0^{(0)}(\Lambda) \hat Q^{(1/2,1/2)}_{ijk} 
  ~\delta_\Lambda(\rvec_{ij})~\delta_\Lambda(\rvec_{jk}),
\end{align}
where $\hat P ^{(I,S)}_{ij}$ and $\hat Q ^{(I,S)}_{ijk}$ are projectors to the $s$-wave two- and three-body isospin-spin ($I,S$) channels, respectively. Upon regularization, all low-energy constants (LECs) $C_0^{(0)}(\Lambda)$, $C_1^{(0)}(\Lambda)$, and $D_0^{(0)}(\Lambda)$ gain specific cutoff dependence and are constrained for each $\Lambda$ to a set of available low-energy data. Here, $C_0^{(0)}$ is fitted to reproduce the experimental spin-singlet neutron-neutron scattering length $a_{nn}^0 = -18.95$~fm \cite{nnSCTlngth1,nnSCTlngth2}, $C_1^{(0)}$ to the deuteron binding energy $B({^2\text{H}}) = 2.2246$~MeV \cite{deuteronBE}, and $D_0^{(0)}$ to the triton binding energy $B({^3\text{H}}) = 8.482$~MeV \cite{tritonBE}.

%=========================== Next-to-leading order

Following Ref.~\cite{SB23}, the NLO \nopieft\ potential,
\begin{align}
  V^{(1)} &=  
  \sum _{i<j} \Big( C_0^{(1)}(\Lambda) \hat P ^{(1,0)}_{ij} 
                 + C_1^{(1)}(\Lambda) \hat P ^{(0,1)}_{ij} \Big) 
   \delta_\Lambda(\rvec_{ij})
   \cr & +
   \sum _{i<j<k} \sum _{\text{cyc}}
   D^{(1)}_0(\Lambda) \hat Q ^{(1/2,1/2)} _{ijk}  
   \delta_\Lambda(\rvec_{ij})\delta_\Lambda(\rvec_{jk})
   \cr & + 
   \sum _{i<j} \Big( C_3^{(1)}(\Lambda) \hat P ^{(1,0)} _{ij} 
                 + C_4^{(1)}(\Lambda) \hat P ^{(0,1)} _{ij}
                 \Big) 
\Big( \delta_\Lambda(\rvec_{ij}) \overrightarrow{\nabla}^2_{ij} 
        + \overleftarrow{\nabla}^2_{ij}  \delta_\Lambda(\rvec_{ij})\Big)  
   \cr & + 
   \sum _{i<j<k<l}  E_0^{(1)}(\Lambda) 
   \hat S ^{(0,0)} _{ijkl}  
   \prod_{ab \in \text{pairs}} \delta_\Lambda(\rvec_{ab})
   \label{eq:V_NLO}
\end{align}
is treated using first-order perturbation theory. At NLO, two-body momentum dependent interaction associated with the $C_3^{(1)}(\Lambda)$ and $C_4^{(1)}(\Lambda)$ LECs is included in both $s$-wave $NN$ channels. The perturbative insertion of the NLO \nopieft\ potential changes the specific $\Lambda$-dependence of the LO LECs required by the renormalization. Consequently, two- and three-body NLO counter-terms with the $C_0^{(1)}(\Lambda)$, $C_1^{(1)}(\Lambda)$, and $D_0^{(1)}(\Lambda)$ LECs, and the same form as the LO potential, need to be introduced at this order. Furthermore, an additional four-body force with the $E_0^{(1)}(\Lambda)$ LEC and four-body projector $\hat S ^{(I,S)} _{ijkl}$ must enter the spatial-symmetric $I,S=(0,0)$ four-body channel to properly renormalize the theory \cite{SB23}.  In total, at NLO six low-energy constants are fitted to reproduce experimental data. These are the three LO data points presented above, and the spin-singlet neutron-neutron effective range $r_{nn}^0 = 2.75$~fm \cite{MNS90}, the spin-triplet neutron-proton effective range from the effective range expansion (ERE) around the deuteron pole $r_{np}^1 = 1.753$~fm \cite{STS95}, and the $^4$He binding energy $B({^4\text{He}}) = 28.3$ MeV \cite{heliumBE}.   

%=================================
\section{Methods} \label{sec:meth}
%=================================

%=========================== Scattering in HO trap & Bush formula
While LO and NLO \nopieft\ LECs are fitted in free space, the elastic $n-{\rm ^4 He}$ scattering is studied utilizing an harmonic oscillator (HO) trap. This method was successfully used in calculations of $s$-wave \cite{SB23} and $p$-wave \cite{CSKL23} few-body scattering, where in the latter it proved to be competitive with the Faddeev-Yakubovsky approach. As a result, we briefly introduce it here.

We consider a HO potential
\begin{equation}
V_{\text{HO}} (\rvec _1, \dots, \rvec _A) = \sum _{i<j}  \frac{1}{2}\frac{m}{N} \omega ^2 \rvec _{ij}^2
\label{eq:V_HO_r}
\end{equation}
in the LO \nopieft\ $N$-body Hamiltonian. Here, $m$ is the nucleon mass, and $\omega$ stands for the HO frequency. From this point on we will use natural units $\hbar = c =1$ and set the nucleon mass parameter $(\hbar c)^2/m = 41.47 \,\rm{MeV \cdot fm}^{2}$.

Solving the ${\rm ^5 He}(J^\pi = 1/2^+)$ system with the HO potential added to the \nopieft\ one gives rise to a spectrum of bound states. States with energies below the $\rm ^3H$~+~$\rm ^2H$ threshold correspond, within our theory up to NLO, solely to the $s$-wave $n$-$^4\text{He}$ scattering inside the HO trap. 
We select $\omega$ in Eq.~(\ref{eq:V_HO_r}) such that the HO trap length $b_{\rm HO} = \sqrt{2/(m \omega)}$ is larger than any other scale in the system - \nopieft\ interaction range $\sim \Lambda^{-1}$, size of the $^4\text{He}$ sub-cluster, and range of the induced $n$-$^4\text{He}$ interaction. Under these assumptions one can match the $n$-$^4\text{He}$ asymptotic part of the trapped ${\rm ^5 He_{1/2^+}}$ wave function with the free-space two-body $n$-$^4\text{He}_{0^+}$ asymptotic behavior. Then, scattering information can be extracted from the matching condition connecting the $s$-wave $\delta^{1/2}_{n{\rm ^4 He}}(k)$ phase shifts to the corresponding bound-state energy spectrum in the trap \cite{BERW98,SLB09}
\begin{equation}
k \cot \left [ \delta^{1/2}_{n{\rm ^4 He}}(k) \right] = - \sqrt{4~\mu~\omega}~\frac{ \Gamma \left( 3/4 - \epsilon_\omega / 2\omega \right) }{ \Gamma \left( 1/4 - \epsilon_\omega / 2\omega \right)}.
\label{eq:busch}
\end{equation}
Here, $k$ stands for the relative $n$-$^4\text{He}$ momentum and $\mu\approx 4m/5$ is the respective reduced mass. $\epsilon_\omega = E_\omega(^5\text{He}_{1/2^+})-E_\omega(^4\text{He}_{0^+})$, entering both $\Gamma$-functions, is the energy of the trapped $^5 \rm{He}_{1/2^+}$ bound state with respect to the $n$-$^4\text{He}$ energy threshold. The threshold position is given by the trapped $^4\text{He}_{0^+}$ ground state energy, $E_\omega(^4\text{He}_{0^+})$.

As the LO \nopieft\ interaction is iterated, the corresponding phase shifts are determined from the calculated $E^\text{(LO)}_\omega(^5 \text{He}_{1/2^+})$ and $E^\text{(LO)}_\omega(^4 \text{He}_{0^+})$ energies by applying Eq.~(\ref{eq:busch}). At NLO we consider the $V^\text{(1)}$ potential within first-order perturbation theory. 
The NLO phase shifts are then extracted from the corrected energies, $E^\text{(NLO)}_\omega(^5 \text{He}_{1/2^+})$ and $E^\text{(NLO)}_\omega(^4 \text{He}_{0^+})$, using Eq.~(\ref{eq:busch}). The scattering length $a_{n {\rm ^4He}}^{1/2}$ and effective range $r_{n {\rm ^4He}}^{1/2}$ are determined by fitting calculated phase shifts with the ERE
\begin{equation}
k \cot \left [ \delta^{1/2}_{n{\rm ^4 He}}(k) \right] \simeq  -\frac{1}{a_{n {\rm ^4He}}^{1/2}} + \frac{1}{2} r_{n {\rm ^4He}}^{1/2}~k^2 + \dots,
\label{eq:eff_range_exp}    
\end{equation}
where the dots denote higher-order terms.

%=========================== Stochastic Variational Method

In order to solve the few-body Sch\"rodinger equation we expand the total wave function $|\Psi \rangle$ into a sum of non-orthogonal correlated-Gaussian basis states \cite{SuzVar98}
\be
\langle \xvec | \Psi \rangle = \sum_{i=1}^M c_i ~ \AS \left [{\rm exp}\left(-\frac{1}{2} \xvec ^T \Avec^i \xvec \right)  \otimes  | \varphi_{S,M_S}^i \rangle \otimes |\xi_{I,M_I}^i \rangle \right ],
\ee
where $\AS$ is the antisymmetrization operator, $\Avec^i$ denotes an $(N-1)\times (N-1)$ symmetric positive-definite matrix of non-linear basis state parameters, and $\xvec$ is the vector of Jacobi coordinates. Energies and variational parameters $c_i$ are obtained as a solution of a generalized eigenvalue problem.

We construct the spin $| \varphi_{S,M_S}^i \rangle$ and isospin $|\xi_{I,M_I}^i \rangle$ parts by a successive couplings of single particle states. The $N$-body spin wave function
\begin{equation}
| \varphi_{S,M_S}^i \rangle = \left |[~[~[~[s_1 \otimes s_2]_{s_{12}^i} \otimes s_3]_{s_{123}^i} \dots]_{s^i_{1 \dots (N-1)}} \otimes s_N]_{S,M_S} \right \rangle
\end{equation}
is determined by a set of intermediate spin quantum numbers $(s_{12}^i,s_{123}^i,\dots,s^i_{1 \dots (N-1)})$. In order to span the whole spin space, we consider all possible intermediate configurations that couple to the total spin $S$. For a five body system with total spin $S=1/2$, there are five different couplings for $S$. We construct the isospin part $|\varphi_{I,M_I}^i \rangle$ in the same manner, obtaining a total of 25 unique isospin-spin configurations to be included in the ${\rm ^5 He}_{1/2^+}$ wave function.

%=========================== Refinement/generation of correlated Gaussian basis states
The extraction of the phase shifts via Eq. \eqref{eq:busch} involves a ratio of $\Gamma$-functions, which is particularly sensitive to $\epsilon_\omega$. This demands an accurate calculation of both LO energy and LO wave function, which enters at NLO via the first-order perturbation theory. We use the Stochastic Variational Method (SVM) \cite{SuzVar98} to select an appropriate combination of non-linear parameters and specific spin-isospin configuration for each basis state. This approach yields sufficiently accurate results for the $^4\text{He}_{0^+}$ system. However, a relatively slow convergence is observed for $^5\text{He}_{1/2^+}$ inside the HO trap. This is not surprising, since one must take into account at the same time two considerably different scales - the tightly bound $^4 \text{He}_{0^+}$ core and the extended $n-^4 \text{He}_{0^+}$ part of the trapped $^5\text{He}_{1/2^+}$ wave function.

In order to circumvent this issue, we first select roughly $M \approx 800$ $^5 \text{He}_{1/2^+}$ basis states via the SVM. In the second step, we follow the procedure of Ref.~\cite{HSA12} and we enhance the SVM-selected $^5 \text{He}_{1/2^+}$ basis with new correlated Gaussians designed to describe the spatially extended part between the $^4\text{He}_{0^+}$ core and one additional neutron. After generating via SVM 300 $^4 \text{He}_{0^+}$ basis states, one extra neutron is coupled to each of the $^4 \text{He}_{0^+}$ core states through an expansion in the $\xvec_4$ Jacobi coordinate, corresponding to the relative $n$-$^4\text{He}$ distance,
\begin{equation}
  \begin{split}
    &\phi^{~i,~n}_{1/2^+}(^5 \text{He}) =\\
    & \AS \left[ \phi^{~i}_{0^+} (^4 \text{He})\otimes 
      ~\left[\text{exp}\left(-\xvec_4^2/(n \beta)^{2}\right) \left|
        \varphi_{1/2}(5) \right> \right]_{1/2}~ \right]_{1/2 M_S}\\
    & \otimes \left[ ~\left| \xi^{i}_{0}(^4 \text{He})\right> \otimes \left|
      \xi_{1/2}(5) \right>~\right]_{1/2~-1/2}.
  \end{split}
  \label{eq:basis_gen}
\end{equation}
Here, $\phi^{~i}_{0^+} (^4 \text{He})$ is the spatial-spin part of the $i$-th $^4 \text{He}_{0^+}$ core basis state with the corresponding isospin part $\xi^{i}_{0}(^4 \text{He})$. $\left| \varphi_{1/2}(5) \right>$ and $\left| \xi_{1/2}(5)\right>$ are the single-neutron spin and isospin wave functions, respectively. An integer number $n$ runs from 1 to 10 and a parameter $\beta$ is varied within a range $\beta \in [0.4; 2.0]$~fm. In total, our procedure yields roughly $M=3800$ basis states, where a certain amount is discarded upon inspection of the linear dependency and numerical stability of our solution.

\begin{figure}
    \centering % 
    \includegraphics[width = \columnwidth]{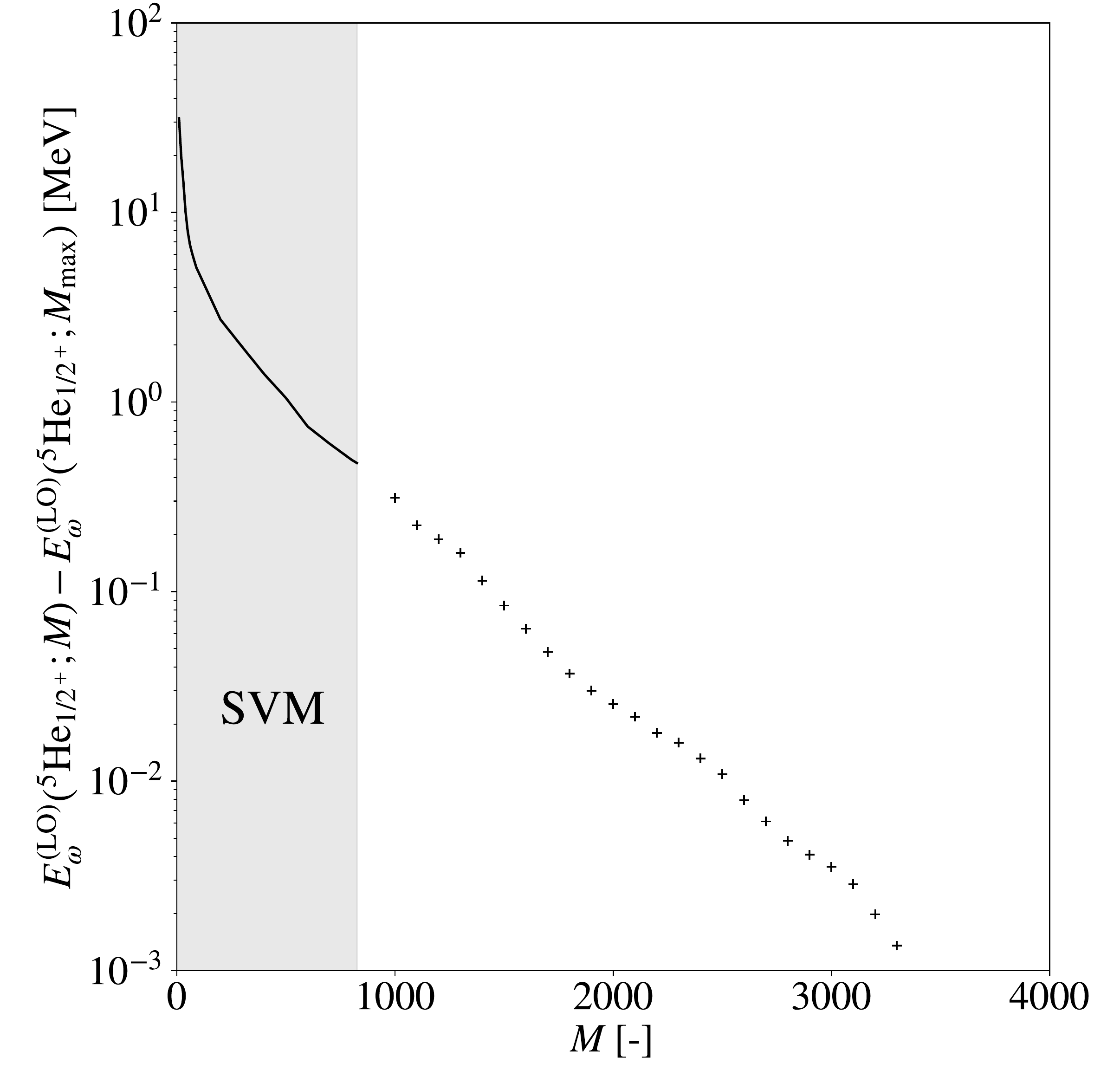}
    \caption{The convergence of $E_\omega^{\text{(LO)}}(^5 \text{He})$ with the number of basis states $M$, relative to the calculation with maximal amount of basis states we considered $M_\text{max}=3522$. 
Presented are energies for momentum cutoff $\Lambda =6~\text{fm}^{-1}$ and $b_\text{HO}= 15~\text{fm}$. Shaded area marks values obtained via the SVM, while the remaining energies were calculated including additional correlated Gaussians generated via Eq.~(\ref{eq:basis_gen}).}    
    \label{fig:convergence}
\end{figure}

\begin{figure*}
    \centering %
    \begin{tabular}{cc}
    \includegraphics[width = \columnwidth]{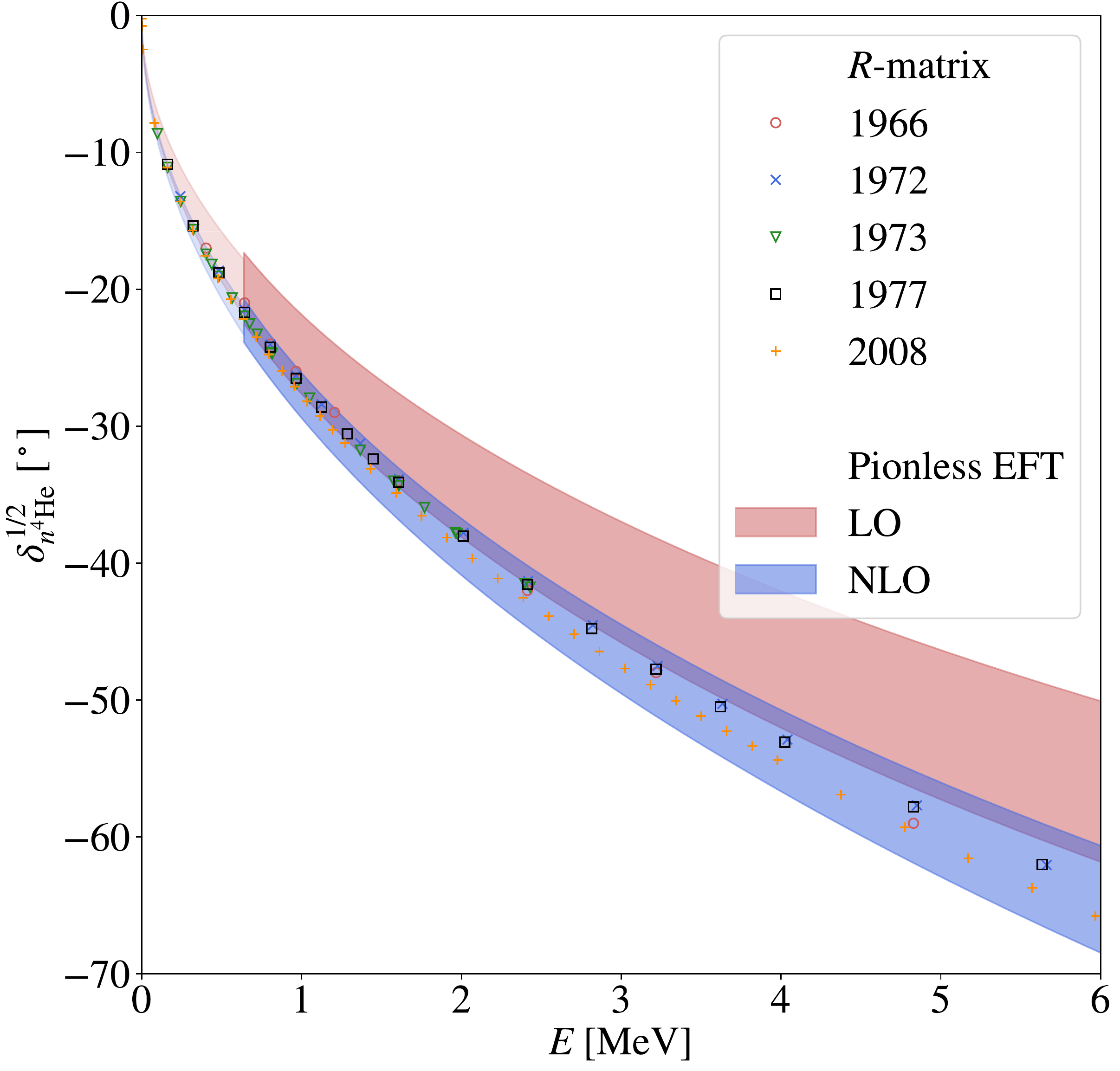} & \includegraphics[width = \columnwidth]{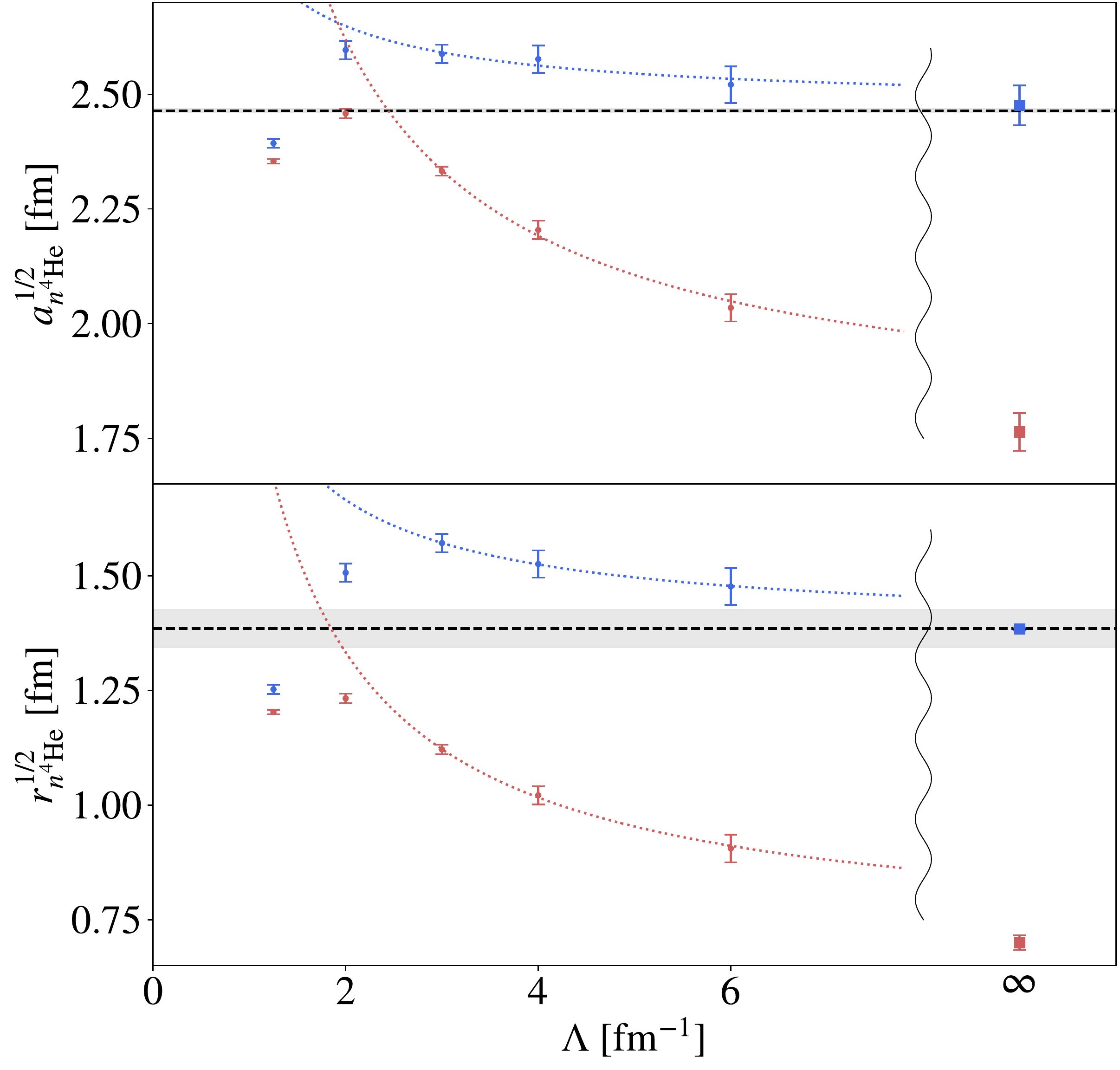}\\
    \end{tabular}
    \caption{Left panel: LO (red) and NLO (blue) \nopieft\ $s$-wave $n$-$^4\text{He}$ phase shifts $\delta^{1/2}_{n^4\text{He}}$ as a function of relative energy $E$. The shaded areas show a spread of corresponding values given by different momentum cutoffs $\Lambda \in [1.25,6.0]~\text{fm}^{-1}$. For $E \lesssim 0.7$~MeV, all phase shifts (displayed in lighter colours) were obtained using ERE, Eq.~(\ref{eq:eff_range_exp}), fitted to our $E \gtrsim 0.7$ HO trap results, Eq.~(\ref{eq:busch}). \nopieft\ predictions are compared to $\delta^{1/2}_{n^4\text{He}}$ obtained in different $R$-matrix studies - 1966 \cite{HB66}, 1972 \cite{SW72}, 1973 \cite{AR73}, 1977 \cite{BF77}, and 2008 \cite{Hale08}. Right panel: Fitted $n$-$^4\text{He}$ scattering length $a^{1/2}_{n^4\text{He}}$ (upper part) and effective range $r^{1/2}_{n^4\text{He}}$ (lower part) as a function of increasing momentum cutoff $\Lambda$. The dotted lines stand for a fit of corresponding LO (red) and NLO (blue) results for $\Lambda \geq 3~\text{fm}^{-1}$ using $f(\Lambda)= f(\infty) + \alpha/\Lambda$. The results for $\Lambda \rightarrow \infty$ are shown as the last entry on the right, given by squares with error bars. \nopieft\ results are compared to $a^{1/2}_{n^4\text{He}}$ and $r^{1/2}_{n^4\text{He}}$ values obtained in 1973 $R$-matrix analysis \cite{AR73} (black dashed lines).}
    \label{fig:phase_shifts_LO_NLO}
\end{figure*}

%================================
\section{Results} \label{sec:res}
%================================

We calculate the $^5 \text{He}_{1/2^+}$ and $^4 \text{He}_{0^+}$ ground states in the HO trap by employing the LO \nopieft\ potential for several momentum cutoffs up to $\Lambda = 6$~$\text{fm}^{-1}$. The harmonic oscillator frequencies $\omega$ are selected such that the corresponding trap lengths $b_\text{HO}$ are between 5 and 15~fm. The upper boundary is currently given by numerical limitations of our five-body calculations, since Eq.~(\ref{eq:busch}) requires uncertainties in $E_\omega(^5 \text{He}_{1/2^+})$ below $10^{-3}\,\rm{MeV}$. The lower boundary is chosen to be larger than $n$-$^4\text{He}$ interaction range $a^{1/2}_{n^4\text{He}} \approx r^{1/2}_{n^4\text{He}} \approx R_{n^4\text{He}}$, which can be estimated based on available scattering length and effective range values \cite{GBSC63,REA69,KRBB79,HWAH20,AR73} to be $R_{n^4\text{He}} \approx 2.5\,\rm{fm}$.

%========================= Fig 1 and energy precision ===============
Following the methods described in Sec.~\ref{sec:meth}, we select correlated-Gaussian basis states separately for each $\Lambda$ and $b_{\text{HO}} =$~7.0, 8.5, 10.0, 12.5, and 15.0~fm. In order to access $\delta^{1/2}_{n^4\text{He}}$ phase shifts corresponding to a finer $b_{\text{HO}} \in [5;15]$~fm grid, we diagonalize the trapped LO \nopieft\ Hamiltonian for these trap lengths with the basis generated for $b_{\text{HO}} = 15$~fm and the respective $\Lambda$. We check that such $E_\omega^{\text{(LO)}}(^5 \text{He})$ energies obtained for $b_{\text{HO}} =$~7.0, 8.5, 10.0, 12.5~fm differ from the ones calculated with a fine-tuned basis by less than $\sim 10^{-4}$~MeV.

In Fig.~\ref{fig:convergence}, we show a convergence of $E_\omega^{\text{(LO)}}(^5 \text{He}_{1/2^+})$ energy with an increasing number of basis states $M$ for $\Lambda = 6~\text{fm}^{-1}$ and $b_\text{HO} = 15.0$~fm. Due to a relatively short-ranged \nopieft\ interaction at this high cutoff and, at the same time, rather broad HO trap, such calculation is the most difficult to converge in our study. Employing the basis generation presented in Eq.~(\ref{eq:basis_gen}) ensures rather fast convergence. The ground state energy changes by less than $10^{-2}$~MeV by the addition of the last $\approx 700$ states. After inspecting the convergence patterns of $E_\omega^{\text{(LO)}}(^5 \text{He}_{1/2^+})$ for the other $\Lambda$ and $b_\text{HO}$ values, we estimate that the global uncertainty of our calculated $^5\text{He}_{1/2^+}$ energies does not exceed $10^{-2}$~MeV.

%========================= Phaseshift results and Fig2 left ===============
By using Eq.~(\ref{eq:busch}), we obtain the $s$-wave $n$-$^4\text{He}$ phase shifts for relative energies between $0.7 \lesssim E \lesssim 7.5$~MeV. In the left panel of Fig.~\ref{fig:phase_shifts_LO_NLO}, we show a spread of the corresponding LO (red shaded area) and NLO (blue shaded area) values given by different momentum cutoffs $\Lambda \in [1.25;6]~\text{fm}^{-1}$. In order to address the $n$-$^4\text{He}$ scattering outside of this energy range, we fit $\delta^{1/2}_{n ^4\text{He}}(E)$ using the ERE in Eq.~(\ref{eq:eff_range_exp}) for each calculated $\Lambda$. The lighter bands at the lowest energies show a spread of the LO and NLO phase shifts determined from this fit. We observe that the residual cutoff dependence gets smaller with the higher order because of the narrower NLO band. The inclusion of NLO terms moves the resulting phase shifts slightly downward, where quite satisfying agreement with $R$-matrix studies is achieved \cite{HB66,SW72,AR73,BF77,Hale08}.

%================= Scattering paremeters results and Fig2 right ===============
The fitted $n$-$^4\text{He}$ scattering length $a^{1/2}_{n^4\text{He}}$ and effective range $r^{1/2}_{n^4\text{He}}$ are presented as a function of increasing $\Lambda$ in the right panel of Fig.~\ref{fig:phase_shifts_LO_NLO}. The corresponding NLO results are rather close to the ERE parameters extracted by Arndt and Roper from their 1973 $R$-matrix analysis \cite{AR73} - see the dashed black lines in the same panel. By performing a simple $f(\Lambda) = f(\infty) + \alpha / \Lambda$ fit, we extrapolate our $\Lambda \geq 3~\text{fm}^{-1}$ results to $\Lambda \rightarrow \infty$ . The corresponding $a^{1/2}_{n^4\text{He}}(\infty)$ and $r^{1/2}_{n^4\text{He}}(\infty)$ values are shown as the last entry in the panel using squares with error bars.

%=============== Paragraph about theoretical error estimates ==================
Our work is affected by different sources of error which can be systematically accounted for by estimating their magnitudes. The first error emerges due to the variational nature of our calculation and it enters through the $\epsilon_\omega$ energies in Eq.~(\ref{eq:busch}). Since it is estimated to be less than $\delta \epsilon_\omega \lesssim 10^{-2}$~MeV, we propagate it into our phase-shift bands shown in the left panel of Fig.~\ref{fig:phase_shifts_LO_NLO} and into the scattering parameter error bars in the right panel of the same figure. Another source of error might be because of an insufficient separation between the $b_\text{HO}$ and $R_{n^4\text{He}}$ scales. We tried to apply the method presented in Ref.~\cite{Zhang20}, which suggests a systematic removal of residual HO trap contribution, but we observed only a negligible effect on our results. 

In our study, the main source of uncertainty is introduced by the truncation of the \nopieft\ expansion at LO or at NLO, i.e. by the theoretical error. This contribution is estimated by inspecting the residual cutoff dependence \cite{Gri20} of the calculated ERE parameters depicted in the right panel of Fig.~\ref{fig:phase_shifts_LO_NLO}. At the given order, for each $a^{1/2}_{n^4\text{He}}$ and $r^{1/2}_{n^4\text{He}}$, we deduce this error from the spread of the corresponding values calculated at cutoffs much larger than the break-up scale of the theory $\Lambda >> m_{\pi}$. More specifically, we consider results for $\Lambda \in [3.0;\infty]~\text{fm}^{-1}$.

Our final $a^{1/2}_{n^4\text{He}}$ and $r^{1/2}_{n^4\text{He}}$ predictions are given as extrapolated values for $\Lambda \rightarrow \infty$ accompanied by two separate errors, numerical and theoretical, 
\be
\begin{split}
    \text{LO}~~  &:~~ a^{1/2}_{n ^4\text{He}}=1.76(4\ \text{num.})~(62\ \text{theor.})~{\rm fm},\\[4pt]
    &~~~~~r^{1/2}_{n ^4\text{He}}=0.70(2\ \text{num.})~(45\ \text{theor.})~{\rm fm},\\[6pt]
    \text{NLO} &:~~ a^{1/2}_{n ^4\text{He}}=2.47(4\ \text{num.})~(17\ \text{theor.})~{\rm fm},\\[4pt]
    &~~~~~r^{1/2}_{n ^4\text{He}}=1.384(3\ \text{num.})~(211\ \text{theor.})~{\rm fm}.\\
\end{split}
\nonumber
\ee
As expected, theoretical errors decrease with the inclusion of NLO terms. Furthermore, they are larger by an order of magnitude than the numerical ones.

%======== Comparison to a0 and r0 of experiments and theor res. in Fig3 =======
A comparison of our final $a^{1/2}_{n ^4\text{He}}$ and $r^{1/2}_{n ^4\text{He}}$ results to experimental and other theoretical works is presented in Fig.~\ref{fig:a0_r0}. The predicted scattering length can be directly compared to the available data provided by transmission \cite{GBSC63,KRBB79} and neutron interferometry \cite{REA69,HWAH20} measurements. Considering the $R$-matrix studies, only the 1973 analysis \cite{AR73} provides explicit result for both $ a^{1/2}_{n ^4\text{He}}$ and $r^{1/2}_{n ^4\text{He}}$. We fit the available phase shifts up to $E=8$~MeV of the remaining $R$-matrix works \cite{HB66,SW72,BF77,Hale08} using the ERE in Eq.~(\ref{eq:eff_range_exp}), thus extracting the scattering length and the effective range. In order to assess the stability of the fit, we vary the maximal energy between $4\leq E \leq 8$~MeV. Then, this uncertainty is inserted in the corresponding $a^{1/2}_{n ^4\text{He}}$ and $r^{1/2}_{n ^4\text{He}}$ error bars.

Regarding the microscopic few-body calculations, a Green's function Monte Carlo (GFMC) study was performed in Ref.~\cite{NPWC07} using AV18 $NN$ potential \cite{AV18} with and without Urbana IX (UIX) \cite{UIX} or Illinois-2 (IL2) \cite{IL2} three-body $NNN$ force. It was reported that all calculations are consistent with $a^{1/2}_{n ^4\text{He}}=2.4$~fm. Consequently, we mark this result in the left panel of Fig.~\ref{fig:a0_r0} as AV18(+UIX/IL2). The $n$-$^4\text{He}$ elastic scattering was further studied within the five-body FY formalism using different models of nuclear interactions \cite{Lazauskas18,Lazauskas20,LC20}. We display the respective results for AV18, AV18+UIX, Reid93 \cite{SKTS94}, Malfliet-Tjon (MT) I-III \cite{MT69}, INOY04 \cite{Doleschall04}, and $\chi$EFT(N$^3$LO) $NN$ interaction \cite{EM03} with and without $V_3$(N$^2$LO) $NNN$ force \cite{MKRS12}. In the FY studies, the $a^{1/2}_{n ^4\text{He}}$ scattering length values with the corresponding numerical error were given directly by the authors. Furthermore, for several nuclear interactions AV18, MT~I-III, INOY04, $\chi$EFT(N$^3$LO), $\chi$EFT(N$^3$LO) + $V_3$(N$^2$LO) the low-energy $s$-wave phase shifts were provided as well. In these cases, in order to obtain the effective ranges $r^{1/2}_{n ^4\text{He}}$, we fit the calculated phase shifts with the ERE, Eq.~(\ref{eq:eff_range_exp}), fixing at the same time the corresponding scattering length to the published $a^{1/2}_{n ^4\text{He}}$ value. As for $R$-matrix studies, we assess the stability of the fit by varying the maximal energy of the considered phase shifts. This uncertainty contributes to the $r^{1/2}_{n ^4\text{He}}$ error bars in the left panel of Fig.~\ref{fig:a0_r0}. Low-energy $s$-wave $n$-$^4\text{He}$ phase shifts were also obtained in SS-HORSE calculations \cite{SMMV16,SMMM18} using JISP16 \cite{Jisp16} and Daejon16 \cite{Daejon16} interactions, and in NCSMC calculation \cite{ZSNG20} with $\chi$EFT(N$^2$LO$_\text{opt}$) $NN$ force \cite{EBFH13}. The ERE parameters $a^{1/2}_{n ^4\text{He}}$ and $r^{1/2}_{n ^4\text{He}}$ are extracted just like in the case of the aforementioned $R$-matrix studies. It is to be noted that the elastic $n$-$^4\text{He}$ scattering was further addressed in Refs.~\cite{QN08,QN09,NRQ10,HLNQ13,NQHRR16,KQQW20,MLDE22}. However, corresponding phase-shift values are not accessible in order to perform a more detailed comparison. 

Our \nopieft\ results are depicted in Fig.~\ref{fig:a0_r0} with two sets of error bars - the smaller ones denote the corresponding numerical error, while the larger error bars represent the total error inflated by the estimated LO or NLO theoretical uncertainty. We emphasize that the results of other microscopic calculations are presented in the figure only with their respective numerical error. In fact, the estimation of theoretical uncertainty, which is a difficult task to address, is a strong feature of this work.       

%====================================
\section{Conclusions} \label{sec:con}
%====================================

We provided the first five-body calculation of low-energy $s$-wave $n$-$^4\text{He}$ elastic scattering within the $\slashed{\pi}$EFT framework, including effective range corrections perturbatively. We constrained the theory to just six well-established experimental results and we applied the harmonic oscillator trap technique to extract the corresponding phase shifts at LO and NLO.

%errors and standing with experiments and other theoretical results
Special care was taken to estimate the numerical as well as theoretical errors, where the latter was accessed based on the residual cutoff dependence of our results. We find that the main source of uncertainty is the theoretical error, which decreases upon the inclusion of NLO terms. This leaves our $a^{1/2}_{n ^4\text{He}}$ and $r^{1/2}_{n ^4\text{He}}$ predictions at NLO with an estimated theoretical uncertainty of order of $7\%$ and $15\%$, respectively. 

\begin{figure*}
    \centering % 
    \begin{tabular}{cc}
    \includegraphics[width = 0.887\columnwidth]{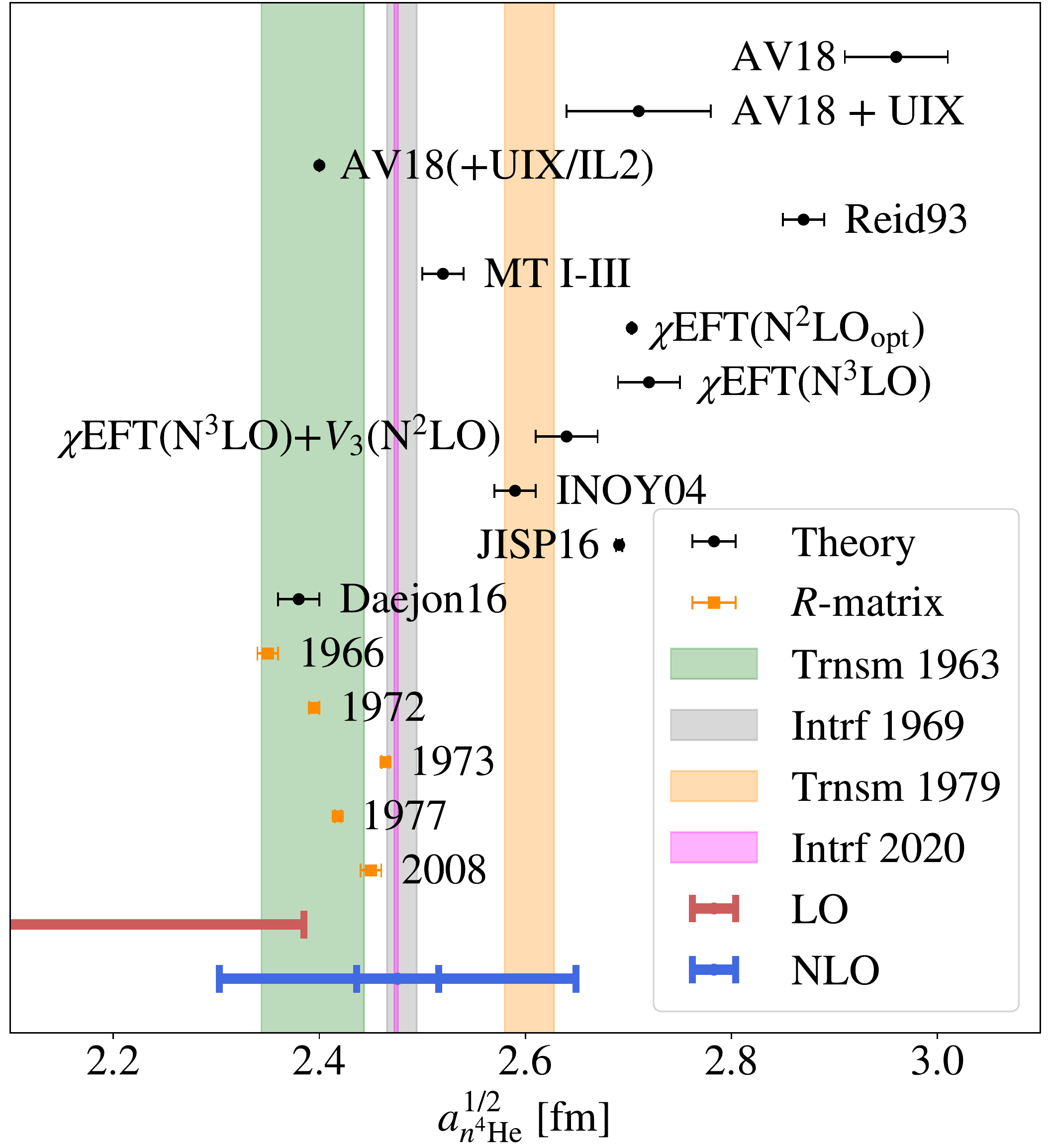} & \includegraphics[width = 0.887\columnwidth]{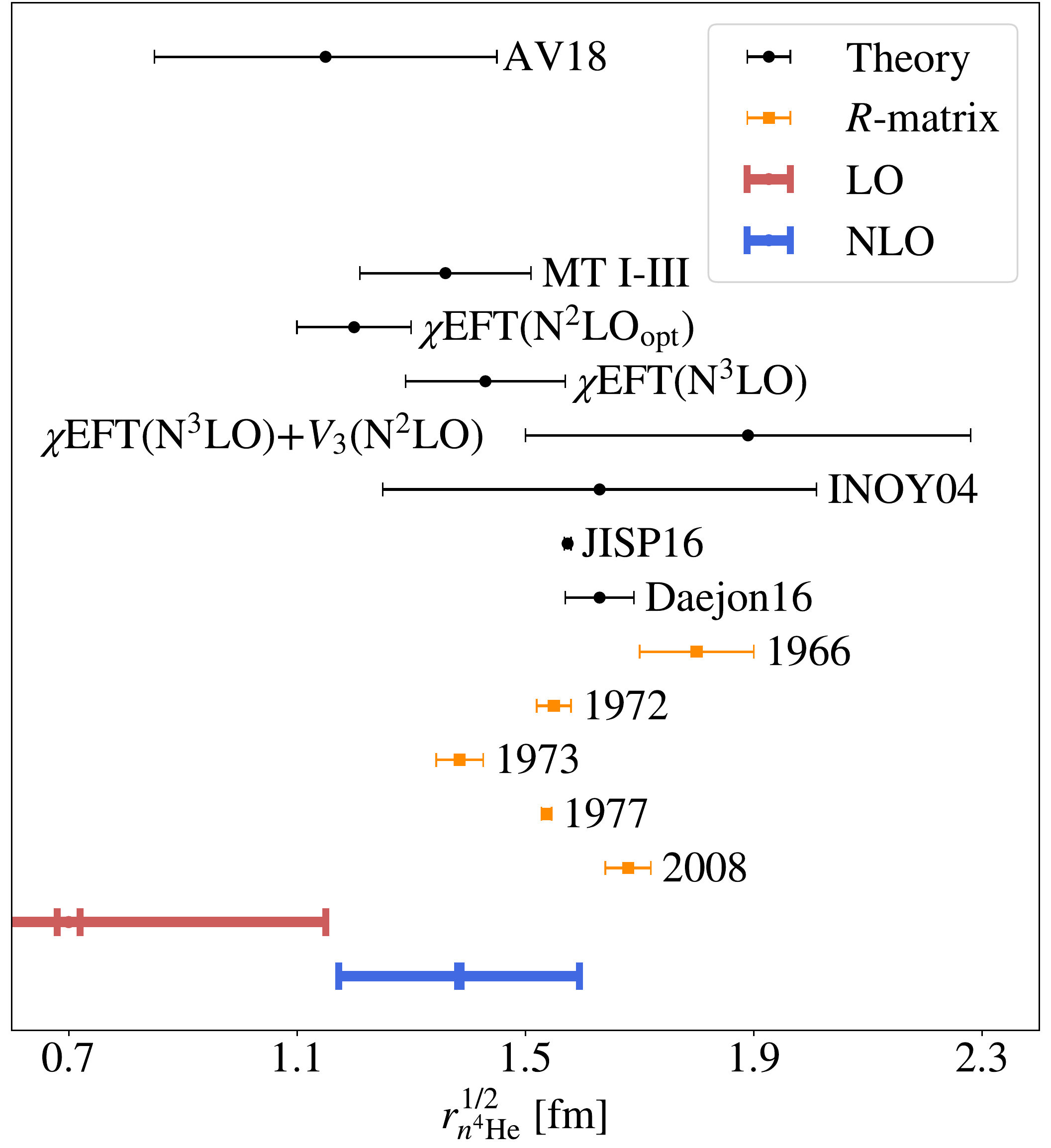}\\
    \end{tabular}
    \caption{ Left panel: Final LO (red) and NLO (blue) \nopieft\ results for $n$-$^4\text{He}$ scattering length $a^{1/2}_{n^4\text{He}}$, obtained in this work, compared to data from transmission measurements (Trnsm) 1963 \cite{GBSC63} and 1979 \cite{KRBB79}, neutron interferometry (Intrf) 1969 \cite{REA69} and 2020 \cite{HWAH20}, and $R$-matrix analysis 1966 \cite{HB66}, 1972 \cite{SW72}, 1973 \cite{AR73}, 1977 \cite{BF77}, 2008 \cite{Hale08}. Our results are further compared to predictions of various few-body calculations considering different nuclear interactions - AV18(+UIX/IL2) \cite{NPWC07}; AV18, AV18+UIX, Reid93, MT I-III, $\chi$EFT(N$^3$LO), $\chi$EFT(N$^3$LO) + $V_3$(N$^2$LO), INOY04 \cite{Lazauskas18,Lazauskas20,LC20}; $\chi$EFT(N$^2$LO$_\text{opt}$) \cite{ZSNG20}; JISP16, Daejon16 \cite{SMMV16,SMMM18}. Right panel: The same as on the left but for $n$-$^4\text{He}$ effective range $r^{1/2}_{n^4\text{He}}$. For AV18(+IL2/UIX), AV18+UIX, and Reid93 only scattering lengths are available. The smaller error bars in our \nopieft\ results show numerical uncertainty only, while the total error bar represents the contribution from both the numerical and estimated theoretical error. For details on errors corresponding to $R$-matrix values and results provided by different few-body calculations see the text.}
    \label{fig:a0_r0}
\end{figure*}

Apparently, $n$-$^4\text{He}$ scattering length is fairly well constrained by the accurate 2020 neutron interferometry measurement \cite{HWAH20} which is in agreement with the earlier interferometry datum \cite{REA69}. These results are close to the value given by neutron transmission \cite{GBSC63} or suggested by $R$-matrix works \cite{HB66,SW72,AR73,BF77,Hale08}. As can be seen in the left panel of Fig.~\ref{fig:a0_r0}, most microscopic predictions tend to yield slightly larger $a^{1/2}_{n ^4\text{He}}$. Between the depicted values, only GFMC AV18(+UIX/IL2), FY MT~I-III, SS-HORSE Daejon16, and our NLO \nopieft\ calculations predict the scattering length close to or in agreement with the most recent experimental result. We emphasize that for $a^{1/2}_{n ^4\text{He}}$ there is a non-negligible  difference between FY and GFMC studies using the same AV18 or AV18+UIX nuclear potentials. It was suggested in Ref.~\cite{LC20} that the main reason might be a lack of accuracy in GFMC calculations. For the effective range, the possible $r^{1/2}_{n ^4\text{He}}$ values given in the right panel of Fig.~\ref{fig:a0_r0} occupy a relatively broad interval. Some results are further burdened by large uncertainties propagated from the ERE fit to the corresponding phase shifts. We can deduce that our NLO \nopieft\ prediction is in agreement with most $R$-matrix values or results of microscopic calculations. However, it is difficult to conclude a more refined statement.

%4-body and clusterization
We demonstrated that the inclusion of NLO four-body force in isospin-spin $I,S=(0,0)$ four-body channel leads not only to renormalization-group invariant results but also to the prediction of the $a^{1/2}_{n ^4\text{He}}$ scattering length in agreement with the experimental measurements. This result surpasses the \nopieft\ study in Ref.~\cite{SB23} where the invariance at NLO was for the first time shown by studying the $s$-wave $N-3N$ scattering in the same channel. However, not considering the Coulomb interaction in that work did not allow a direct comparison of the calculated scattering to experimental data.

%higher partial waves
Future studies should address the few-body scattering in higher partial waves. Starting with the $p$-wave, there is a series of experimentally observed low-laying resonances in $^4\text{H}$ and $^5 \text{He}$ nuclear systems which strongly affect the corresponding phase shifts. In fact, a recent calculation of $^4\text{H}$ ($J^\pi = 1^-$) resonance \cite{CSKL23} using LO SU(4)-symmetric \nopieft\ demonstrated numerically that the corresponding resonance pole seems to be cutoff-stable as $\Lambda \rightarrow \infty$. Consequently, it is highly topical to check whether perturbative inclusion of higher-order terms predicts the corresponding resonance poles closer to or in agreement with experimental data. Such information would provide a strong theoretical insight into unbinding of $A>4$ nuclear systems observed so far within LO \nopieft\ \cite{SBK07,CLPR17,DCKG20,SCKM21}.      

%=========================
\section*{Acknowledgement}
%=========================

The work of M. Bagnarol, N. Barnea, and M. Sch\"{a}fer was supported by the Israel Science Foundation grant 1086/21 and by the European Union’s Horizon
2020 research and innovation program under grant agreement No 824093. The work of M.S. was also supported by the Czech Science Foundation GA\v{C}R grant 19-19640S. This research was supported in part by the National Science Foundation under Grant No. NSF PHY-1748958. 

%==========================


\begin{thebibliography}{99}
%==========================

%================== experimental scattering length references ======================
\bibitem{GBSC63} R. Genin, H. Beil, C. Signarbieux, P. Carlos, R. Joly, and M. Ribrag,
  D\'{e}termination des sections efficaces d'absorption et de diffusion des gaz rares pour les neutrons thermiques,
  Le Journal de Physique et le Radium 24 (1963) 21.

\bibitem{REA69} D. C. Rorer, B. M. Ecker, and R. \"{O}. Aky\"{u}z,
  Slow-neutron cross sections of He, Ne, Ar and Kr, 
  Nucl. Phys. A 133 (1969) 410.

\bibitem{KRBB79} H. Kaiser, H. Rauch, G. Badurek, W. Bauspiess, and U. Bonse,
  Measurement of Coherent Neutron Scattering Lengths of Gases
  Zeit. Phys. A291 (1979) 231.

\bibitem{HWAH20} R. Haun, F. E. Wietfeldt, M. Arif, M. G. Huber, T. C. Black, B. Heacock, D. A. Pushin, and C. B. Shahi,
  Precision Measurement of the Neutron Scattering Length of $^4$He Using Neutron Interferometry,
  Phys. Rev. Lett. 124 (2020) 012501.

%========================== R-matrix references =================================

\bibitem{HB66} B. Hoop Jr. and H. H. Barschall, 
  Scattering of neutrons by $\alpha$-particles,
  Nucl. Phys. 83 (1966) 65.

\bibitem{SW72} Th. Stammbach and R. L. Walter,
  $R$-matrix formulation and phase shifts for n-$^4$He and p-$^4$He scattering for energies up to 20 MeV,
  Nucl. Phys. A 180 (1972) 225.

\bibitem{AR73} R. A. Arndt and L. D. Roper,
  Nucleon-alpha elastic scattering analyses: (II). 0 to 21 MeV energy-dependent n-$\alpha$ analysis,
  Nucl. Phys. A 209 (1973) 447.

\bibitem{BF77} J. E. Bond and F. W. K. Firk,
  Determination of $R$-function and physical-state parameters for n-$^4$He elastic scattering below 21 MeV,
  Nucl. Phys. A 287 (1977) 317.

\bibitem{Hale08} G. M. Hale (private communication).

%========================== few-body calculations references =================================

%Monte Carlo 
\bibitem{CSK87} J. Carlson, K. E. Schmidt, and M. H. Kalos,
  Microscopic calculations of $^5$He with realistic interactions,
  Phys. Rev. C 36 (1987) 27.

\bibitem{NPWC07} K. M. Nollett, S. C. Pieper, R. B. Wiringa, J. Carlson, and G. M. Hale,
  Quantum Monte Carlo Calculations of Neutron-$\alpha$ Scattering,
  Phys. Rev. Lett. 99 (2007) 022502.

%Faddeev-Yakubovsky
\bibitem{Lazauskas18} R. Lazauskas,
  Solution of the $n$-$^4$He elastic scattering problem using the Faddeev-Yakubovsky equations,
  Phys. Rev. C 97 (2018) 044002.

\bibitem{Lazauskas20} R. Lazauskas,
  Recent Developments in Solving the Few-Particle Scattering Problem by the Solution of The Faddeev-Yakubovsky Equations,
  Recent Progress in Few-Body Physics, FB22 2018, Springer Proceedings in Physics 238 (2020) Springer.

\bibitem{LC20} R. Lazauskas and J. Carbonell,
  Description of Four- and Five-Nucleon Systems by Solving Faddeev-Yakubovsky Equations in Configuration Space,
  Front. Phys. 7 (2020) 251.

%Series of \chiEFT works - Quaglioni and Navratil
\bibitem{QN08} \emph{Ab Initio} Many-Body Calculations of $n$-$^3$H, $n$-$^4$He, $p$-$^{3,4}$He, and $n$-$^10$Be Scattering
  S. Quaglioni and P. Navr\'{a}til,
  Phys. Rev. Lett. 101 (2008) 092501.

\bibitem{QN09} S. Quaglioni and P. Navr\'{a}til,
  \emph{Ab initio} many-body calculations of nucleon-nucleus scattering,
  Phys. Rev. C 79 (2009) 044606.

\bibitem{NRQ10} P. Navr\'{a}til, R. Roth, and S. Quaglioni,
  \emph{Ab initio} many-body calculations of nucleon scattering on $^4$He, $^7$Li, $^7$Be, $^{12}$C, and $^{16}$O,
  Phys. Rev. C 82 (2010) 034609.

\bibitem{HLNQ13} G. Hupin, J. Langhammer, P. Navr\'{a}til, S. Quaglioni, A. Calci, and R. Roth,
  \emph{Ab initio} many-body calculations of nucleon-$^4$He scattering with three-nucleon forces,
  Phys. Rev. C 88 (2013) 054622.

\bibitem{NQHRR16} P. Navr\'{a}til, S. Quaglioni, G. Hupin, C. Romero-Redondo, and A. Calci,
  Unified \emph{ab initio} approaches to nuclear structure and reactions,
  Phys. Scr. 91 (2016) 053002.

\bibitem{ZSNG20} X. Zhang, S. R. Stroberg, P. Navr\'{a}til, Ch. Gwak, J. A. Melendez, R. J. Furnstahl, and J. D. Holt,
  \emph{Ab Initio} Calculations of Low-Energy Nuclear Scattering Using Confining Potential Traps,
  Phys. Rev. Lett. 125 (2020) 112503.

\bibitem{KQQW20} K. Kravvaris, K. R. Quinlan, S. Quaglioni, K. A. Wendt, and P. Navr\'{a}til,
  Quantifying uncertainties in neutron-$\alpha$ scattering with chiral nucleon-nucleon and three-nucleon forces,
  Phys. Rev. C 102 (2020) 024616.

%SS-HORSE
\bibitem{SMMV16} A. M. Shirokov, A. I. Mazur, I. A. Mazur, and J. P. Vary,
  Shell model states in the continuum,
  Phys. Rev. C 94 (2016) 064320; Erratum Phys. Rev. C 98 (2018) 039901.

\bibitem{SMMM18} A. M. Shirokov, A. I. Mazur, I. A. Mazur, E. A. Mazur, I. J. Shin, Y. Kim, L. D. Blokhintsev, and J. P. Vary,
  Nucleon-$\alpha$ scattering and resonances in $^5$He and $^5$Li with JISP16 and Daejeon16 $NN$ interactions,
  Phys. Rev. C 98 (2018) 044624.

%SA-NCSM with RGM
\bibitem{MLDE22} A. Mercenne, K. D. Launey, T. Dytrych, J. E. Escher, S. Quaglioni, G. H. Sargsyan, D. Langr, J. P. Draayer,
  Efficacy of the symmetry-adapted basis for \emph{ab initio} nucleon-nucleus interactions for light- and intermediate-mass nuclei,
  Comp. Phys. Comm. 280 (2022) 108476.

%\chEFTN3LO
\bibitem{EM03} D. R. Entem and R. Machleidt,
  Accurate charge-dependent nucleon-nucleon potential at fourth order of chiral perturbation theory,
  Phys. Rev. C 68 (2003) 041001(R).

%\chEFTN4LO
\bibitem{EMN17} D. R. Entem, R. Machleidt, and Y. Nosyk,
  High-quality two-nucleon potentials up to fifth order of the chiral expansion,
  Phys. Rev. C 96 (2017) 024004.

%========================== pionless EFT references =================================
\bibitem{HKK20} H.-W. Hammer, S. K\"onig, and U. van Kolck,
  Nuclear effective field theory: Status and perspectives,
  Rev. Mod. Phys. 92 (2020) 025004.

\bibitem{Kolck99} U. van Kolck,
  Effective field theory of short-range forces,
  Nucl. Phys. A 645 (1999) 273.

%three-body force
\bibitem{BedHamKol99} P. F. Bedaque, H.-W. Hammer, and U. van Kolck, 
  Renormalization of the Three-Body System with Short-Range Interactions,
  Phys. Rev. Lett. 82 (1999) 463;
  The three boson system with short-range interactions, 
  Nucl. Phys. A 646 (1999) 444;
  Effective theory of the triton,
  Nucl. Phys. A 676 (2000) 357.

%A <=4 NLO scattering work (four-body)
\bibitem{SB23} M. Sch\"{a}fer and B. Bazak,
  Few-nucleon scattering in pionless effective field theory,
  Phys. Rev. C 107 (2023) 064001.

\bibitem{BazKirKon19} B. Bazak, J. Kirscher, S. K\"onig, M. Pav\'on Valderrama, N. Barnea, and U. van Kolck,
  Four-Body Scale in Universal Few-Boson Systems, 
  Phys. Rev. Lett. 122 (2019) 143001.

%momentum dependent three-body
\bibitem{BedRupGri03} P. F. Bedaque, G. Rupak, H. W. Grießhammer, and H.-W.Hammer,
  Low energy expansion in the three body system to all orders and the triton channel,
  Nucl. Phys. A 714 (2003) 589.
  
\bibitem{Gri04} H. W. Grießhammer,
  Improved convergence in the three-nucleon system at very low energies,
  Nucl. Phys. A 774 (2004) 192.

\bibitem{Gri05} H.W. Grießhammer, 
  Na\"{i}ve dimensional analysis for three-body forces without pions,
  Nucl. Phys. A 760 (2005) 110.

%n-d works
\bibitem{BedKol98} P. F. Bedaque and U. van Kolck, 
  Nucleon-deuteron scattering from an effective field theory, 
  Phys. Lett. B 428 (1998) 221.

\bibitem{BedHamKol98} P. F. Bedaque, H.-W. Hammer, and U. van Kolck,
  Effective theory for neutron-deuteron scattering: Energy dependence
  Phys. Rev. C 58 (1998) R641(R).

\bibitem{GabBedGri00} F. Gabbiani, P. F. Bedaque, and H. W. Grießhammer,
  Higher partial waves in an effective field theory approach to scattering,
  Nucl. Phys. A 675 (2000) 601.
  
\bibitem{BedGri00} P. F. Bedaque and H. W. Grießhammer,
  Quartet S wave neutron deuteron scattering in effective field theory,
  Nucl. Phys. A 671 (2000) 357.

\bibitem{HamMeh01} H.-W. Hammer and T. Mehen
  Range corrections to doublet S-wave neutron–deuteron scattering,
  Phys. Lett. B 516 (2001) 353.

\bibitem{Van13} J. Vanasse, 
  Fully perturbative calculation of nd scattering to next-to-next-to-leading order,
  Phys. Rev. C 88 (2013) 044001.

\bibitem{MarSprVan16} A. Margaryan, R. P. Springer, and J. Vanasse,
  nd scattering and the Ay puzzle to next-to-next-to-next-to-leading order,
  Phys. Rev. C 93 (2016) 054001.

\bibitem{Kon17} S. K\"onig,
  Second-order perturbation theory for $^3$He and pd scattering in pionless EFT,
  J. Phys. G: Nucl. Part. Phys. 44 (2017) 064007.

\bibitem{RupVagHig19} G. Rupak, A. Vaghani, R. Higa, and U. van Kolck,
  Fate of the neutron–deuteron virtual state as an Efimov level,
  Phys. Lett. B 791 (2019) 414.

%n-3H/n-3He works
\bibitem{Kir13} J. Kirscher,
  Zero-energy neutron–triton and proton–Helium-3 scattering with \nopieft,
  Phys. Lett. B 721 (2013) 335.
  
\bibitem{KirGriShu10} J. Kirscher, H. W. Grießhammer, D. Shukla, and H. M. Hofmann,
  Universal correlations in pion-less EFT with the resonating group method: Three and four nucleons,
  Eur. Phys. J A 44 (2010) 239.

%n-4He halo EFT study
\bibitem{BHK02} C.A. Bertulani, H.-W. Hammer, and U. van Kolck,
  Effective field theory for halo nuclei: shallow p-wave states,
  Nucl. Phys. A 712 (2002) 37.

%========================== pionless EFT LO & NLO constraints references =====================

%spin-singlet nn scattering length
\bibitem{nnSCTlngth1} D. E. Gonz\'{a}lez Trotter et al., 
  Neutron-deuteron breakup experiment at $E_n =13 MeV$: Determination of the $^1S_0$ neutron-neutron scattering length $a_{nn}$,
  Phys. Rev. C 73 (2006) 034001.

\bibitem{nnSCTlngth2} Q. Chen et al.,
  Measurement of the neutron-neutron scattering length using the $\pi^- d$ capture reaction,
  Phys. Rev. C 77 (2008) 054002.
  
%deuteron binding energy
\bibitem{deuteronBE} C. Van Der Leun and C. Alderliesten,
  The deuteron binding energy,
  Nucl. Phys. A 380 (1982) 261.

%triton binding energy
\bibitem{tritonBE} J. E. Purcell, J. H. Kelley, E. Kwan, C. G. Sheu, and H. R. Weller,
  Energy levels of light nuclei $A=3$,
  Nucl. Phys. A 848 (2010) 1.

%spin-singlet nn effective range
\bibitem{MNS90} G. A. Miller, B. M. K. Nefkens, and I. \v{S}laus,
  Charge symmetry, quarks and mesons,
  Phys. Rep. 194 (1990) 1.

%spin-triplet np effective range (ERE around deuteron pole)
\bibitem{STS95} J. J. de Swart, C. P. F. Terheggen, and V. G. J. Stoks,
  The Low-Energy Neutron-Proton Scattering Parameters and the Deuteron,
  arXiv:nucl-th/9509032 (1995).

%4He binding energy
\bibitem{heliumBE} D. R. Tilley, H. R. Weller, and G. M. Hale,
  Energy levels of light nuclei $A=4$,
  Nucl. Phys. A 541 (1992) 1.

%========================== methods section references =================================

%4H resonance work
\bibitem{CSKL23} L. Contessi, M. Sch\"{a}fer, J. Kirscher, R. Lazauskas, and J. Carbonell,
  Emergence of $^4$H $J^\pi = -1$ resonance in contact theories,
  Phys. Lett. B 840 (2023) 137840.

% trap reference #1
\bibitem{BERW98} T. Busch, B. G. Englert, K. Rzazewski, and M. Wilkens,
  Two cold atoms in a harmonic trap,
  Found. Phys. 28 (1998) 549.

% trap reference #2
\bibitem{SLB09} A. Suzuki, Y. Liang, and R. K. Bhaduri,
  Two-atom energy spectrum in a harmonic trap near a Feshbach resonance at higher partial waves,
  Phys. Rev. A 80 (2009) 033601.

% SVM book
\bibitem{SuzVar98} Y. Suzuki and K. Varga,
  \emph{Stochastic Variational Approach to Quantum-Mechanical Few-Body Problems},
  (Springer, Berlin, 1998).

% 4He photoabsorption work 
\bibitem{HSA12} W. Horiuchi, Y. Suzuki, and K. Arai,
  \emph{Ab initio} study of the photoabsorption of $^4$He,
  Phys. Rev. C 85 (2012) 054002.

%============================== results section =========================================

%Zhang HO trap removal
\bibitem{Zhang20} X. Zhang,
  Extracting free-space observables from trapped interacting clusters,
  Phys. Rev. C 101 (2020) 051602(R).

%residual cutoff dependence
\bibitem{Gri20} H. W. Grießhammer, 
  A consistency test of EFT power countings from residual cutoff dependence. 
  Eur. Phys. J. A 56 (2020) 118.

%AV18
\bibitem{AV18} R. B. Wiringa, V. G. J. Stoks, and R. Schiavilla,
  Accurate nucleon-nucleon potential with charge-independence breaking,
  Phys. Rev. C 51 (1995) 38.

%UIX
\bibitem{UIX} B. S. Pudliner, V. R. Pandharipande, J. Carlson, and R. B. Wiringa, 
  Quantum Monte Carlo Calculations of $A \leq 6$ Nuclei,
  Phys. Rev. Lett. 74 (1995) 4396.

%IL2
\bibitem{IL2} Steven C. Pieper, V. R. Pandharipande, R. B. Wiringa, and J. Carlson, 
  Realistic models of pion-exchange three-nucleon interactions,
  Phys. Rev. C 64 (2001) 014001.

%Reid93
\bibitem{SKTS94} V. G. J. Stoks, R. A. M. Klomp, C. P. F. Terheggen, and J. J. de Swart,
  Construction of high-quality $NN$ potential models,
  Phys. Rev. C 49 (1994) 2950.

%MT I-III
\bibitem{MT69} R. A. Malfliet and J. A. Tjon,
  Solution of the Faddeev equations for the triton problem using local two-particle interactions,
  Nucl. Phys. A 127 (1969) 161.

%INOY04
\bibitem{Doleschall04} P. Doleschall,
  Influence of the short range nonlocal nucleon-nucleon interaction on the elastic $n-d$ scattering: Below 30 MeV,
  Phys. Rev. C 69, 054001 (2004).

%V3(N2LO)
\bibitem{MKRS12} L. E. Marcucci, A. Kievsky, S. Rosati, R. Schiavilla, and M. Viviani,
  Chiral Effective Field Theory Predictions for Muon Capture on Deuteron and $^3$He,
  Phys. Rev. Lett. 108 (2012) 052502; Erratum Phys. Rev. Lett. 121 (2018) 049901.

%JISP16
\bibitem{Jisp16} A. M. Shirokov, J. P. Vary, A. I. Mazur, and T. A. Weber,
  Realistic nuclear Hamiltonian: Ab exitu approach,
  Phys. Lett. B 644 (2007) 33.

%Daejon16
\bibitem{Daejon16} A. M. Shirokov, I. J. Shin, Y. Kim, M. Sosonkina, P. Maris, J. P. Vary,
  N3LO $NN$ interaction adjusted to light nuclei in ab exitu approach,
  Phys. Lett. B 761 (2016) 87.

%\chiEFT(N2LO_opt)
\bibitem{EBFH13} A. Ekstr\"{o}m, G. Baardsen, C. Forss\'{e}n, G. Hagen, M. Hjorth-Jensen, G. R. Jansen, R. Machleidt, W. Nazarewicz, T. Papenbrock, J. Sarich, and S. M. Wild,
  Optimized Chiral Nucleon-Nucleon Interaction at Next-to-Next-to-Leading Order,
  Phys. Rev. Lett. 110 (2013) 192502.

%============================== conclusions section =========================================

%unbound LO \nopieft\ works
\bibitem{SBK07} I. Stetcu, B.R. Barrett, and U. van Kolck,
  No-core shell model in an effective-field-theory framework,
  Phys. Lett. B 653 (2007) 358.

\bibitem{CLPR17} L. Contessi, A. Lovato, F. Pederiva, A. Roggero, J. Kirscher, and U. van Kolck,
  Ground-state properties of $^4$He and $^{16}$O extrapolated from lattice QCD with pionless EFT,
  Phys. Let. B 772 (2017) 839.

\bibitem{DCKG20} W. G. Dawkins, J. Carlson, U. van Kolck, and A. Gezerlis,
  Clustering of Four-Component Unitary Fermions,
  Phys. Rev. Lett. 124 (2020) 143402.

\bibitem{SCKM21} M. Sch\"{a}fer, L. Contessi, J. Kirscher, and J. Mare\v{s},
  Multi-fermion systems with contact theories,
  Phys. Let. B 816 (2021) 136194.

 \end{thebibliography}
\end{document}